\documentclass[a4paper,prd,showkeys,nofootinbib]{revtex4}

\usepackage[T1]{fontenc}
\usepackage[latin1]{inputenc}
\usepackage{graphicx}
\usepackage{amsmath}
\usepackage{amssymb}

\begin{document}

\title{A Generalization of the Goldberg-Sachs Theorem and its Consequences}

\author{Carlos  Batista}
\email{carlosbatistas@df.ufpe.br}
\affiliation{Departamento de Física, Universidade Federal de Pernambuco, 50670-901
Recife - PE, Brazil}

\date{\today}

\begin{abstract}
The Goldberg-Sachs theorem is generalized for all four-dimensional manifolds endowed with torsion-free connection
compatible with the metric, the treatment includes all signatures as well as complex manifolds. It is shown that when the Weyl tensor is algebraically special severe geometric restrictions are imposed. In particular it is demonstrated that the simple self-dual eigenbivectors of the Weyl tensor generate integrable isotropic planes. Another result obtained here is that if the self-dual part of the Weyl tensor vanishes in a Ricci-flat manifold of (2,2) signature the manifold must be Calabi-Yau or symplectic and admits a solution for the source-free Einstein-Maxwell equations.
\end{abstract}

\maketitle

\section{Introduction }
The Petrov classification \cite{Petrov,Stephani} is a form to classify the Weyl tensor in four-dimensional space-times that promoted much progress in general relativity. Besides other contributions it has helped in the search of new exact solutions to Einstein's equation, Kerr metric being the most important example \cite{Kerr}. Particularly important in this process was the Goldberg-Sachs theorem \cite{Goldberg-Sachs}, which associates algebraic constraints in the Weyl tensor to some integrability properties in the space-time. This theorem states that in a Ricci-flat (vacuum) four-dimensional Lorentzian manifold there exists a shear-free null geodesic congruence if, and only if, the Weyl tensor is algebraically special, where such congruences are generated by the so called repeated principal null directions.

Recently the Petrov classification was extended to four-dimensional manifolds of all signatures, as well as complexified manifolds\footnote{In this paper the term "complexified manifold" means a manifold in which the metric can be complex, so that the Weyl tensor is also generally complex. The here called "real manifolds" are the ones with real metric and, consequently, real Weyl tensor. In general the tangent bundle of the real manifolds will be assumed to be complexified. Finally, the term "complex manifold" will mean a manifold that can be covered by complex charts with analytic transition functions, these manifolds are sometimes called Hermitian.}, in an unified treatment based on the action of the Weyl tensor in the bivector bundle \cite{art1}. The intent of the present article is continue this path and explore the generalized version of the Goldberg-Sachs(GS) theorem valid in all signatures \cite{Plebanski2} using the bivector approach and in an unified way, so that the results in real manifolds of any signature follow from the general complex case, by conveniently choosing a real slice. This different form to attack the problem will prove to be valuable because it is full of geometric content. For example, the null eigenbivectors of the Weyl tensor will be shown to generate integrable planes when the Ricci tensor vanishes. The generalized version of the GS theorem in complexified manifolds was investigated before in reference \cite{Plebanski2}, while the Euclidean case was treated in \cite{Broda}, but in both works spinor techniques were used rather than the bivector approach. Some important steps in the direction of this work were also taken in \cite{Robinson Manifolds,Nurwoski2}. Here some progress is made is this subject, in particular it is proved that when the self-dual part of the Weyl tensor vanishes in a Ricci-flat manifold, there is a covariantly constant rank two tensor other than the metric and a solution for the source-free Einstein-Maxwell equations. It is also proved in this article that the Weyl tensor is algebraically special if, and only if, it admits a simple self-dual eigenbivector. It is relevant pointing out that in recent years there has been an active research in order to find a suitable generalization of the GS theorem valid in all dimensions \cite{Dur-Reall,Ortaggio5,HigherGSisotropic1,HigherGSisotropic2}, hopefully the present work can give new insights on these attempts.

Section \ref{bivector/section} provides a short review of the unified algebraic classification scheme for the Weyl tensor in four-dimensional manifolds, there the Weyl tensor is viewed as an operator on the bivector bundle. Next section shows that in Lorentzian signature the repeated principal null directions (PNDs) are in a one to one correspondence with the simple self-dual eigenbivectors of the Weyl tensor. This result serves as the motivation to conjecture that the generalization of the repeated PNDs to the non-Lorentzian manifolds are the simple self-dual eigenbivectors of the Weyl tensor. In section \ref{GSHP/section} the extension of the Goldberg-Sachs theorem to complexified manifolds and real manifolds of all signatures is enunciated. This theorem implies that the simple
self-dual eigenbivectors of the Weyl tensor span integrable planes just as the repeated PNDs are related to the
integrability of shear-free null geodesic congruences, supporting the conjecture of preceding section. Next section provides the interpretations and consequences of the generalized GS theorem. In particular it is proved that when the self-dual part of the Weyl tensor vanishes in complex and Euclidean manifolds the manifold is Calabi-Yau, while in (2,2) signature it can also be symplectic and admits a non-trivial covariantly constant tensor of rank two. Finally, section \ref{Physsection} briefly discuss the physical applicability of achieved results and of the mathematical structure behind them.

\section{Weyl Tensor Classification by Bivectors in All Signatures}\label{bivector/section}
This section will be a quick sum up of the results obtained in \cite{art1}. Let $(M,g_{\mu\nu})$ be a four-dimensional differential manifold endowed with metric $g_{\mu\nu}$. Sometimes $(M,g_{\mu\nu})$ will also denote a complexified Riemannian manifold of complex dimension four, as will be clear in the
context. A skew-symmetric rank two tensor field, $B_{\mu\nu} = - B_{\nu\mu}$, is called a bivector. Denoting the
volume form by $\epsilon_{\mu\nu\rho\sigma}$, the dual of a bivector is defined by:
\begin{equation}\label{dual}
    \widetilde{B}_{\mu\nu} = \frac{1}{2} \epsilon_{\mu\nu\rho\sigma}B^{\rho\sigma}\,.
\end{equation}
It is easy to see that given any two bivectors, $B_{\mu\nu}$ and $F_{\mu\nu}$, we have
 \begin{equation}\label{bx}
    \widetilde{B}_{\mu\nu}F^{\mu\nu} = B_{\mu\nu}\widetilde{F}^{\mu\nu}.
\end{equation}
By means of the contraction properties of the volume form, it is obtained that the double dual of a bivector is a
multiple of it:
\begin{equation}\label{2dual}
    \widetilde{\widetilde{B}_{\mu\nu}} =
    \frac{1}{4}\epsilon_{\mu\nu\rho\sigma}\epsilon^{\rho\sigma\alpha\beta}B_{\alpha\beta} =
    \varepsilon^{2}B_{\mu\nu}\,.
\end{equation}
Where $\varepsilon=1$ in complexified manifolds and in real manifolds of Euclidean or (2,2) signature, while
$\varepsilon=i$ in real Lorentzian manifolds. Equation (\ref{2dual}) enables to split the bivector bundle,
$\mathfrak{B}$, into a direct sum of two spaces of the same dimension:
\begin{equation}\label{oplusc}
\mathfrak{B} = \mathfrak{D}^+ \oplus \,\mathfrak{D}^-,
\end{equation}
$$ \mathfrak{D}^+ = \{Z^+_{\mu\nu} \in  \mathfrak{B} | \widetilde{Z^+}_{\mu\nu}= \varepsilon Z^+_{\mu\nu}\} \; ; \;
 \mathfrak{D}^- = \{Z^-_{\mu\nu} \in  \mathfrak{B} | \widetilde{Z^-}_{\mu\nu}= -\varepsilon Z^-_{\mu\nu}\}. $$
 Where $\mathfrak{D}^+$ is called the bundle of self-dual bivectors, while $\mathfrak{D}^-$ is the bundle of
 anti-self-dual bivectors.

 A bivector $B_{\mu\nu}$ is called simple if it is possible to find two vector fields, $X$ and $Y$, such that
 $B_{\mu\nu}= X_{[\mu} Y_{\nu]}$. This kind of bivector is naturally associated with planes, if $B_{\mu\nu}= X_{[\mu}
 Y_{\nu]}$ the bivector $B_{\mu\nu}$ is said to generate the planes spanned by the vector fields $X$ and $Y$. In four
 dimensions a bivector is simple if, and only if, $B_{\mu\nu}\widetilde{B}^{\mu\nu} = 0$.

 Because of the skew-symmetry in the first and second pairs of Weyl tensor indices it is natural to define
 \begin{equation}\label{Weyldualdefinition}
    C_{\mu\nu\widetilde{\rho\sigma}} = \frac{1}{2} \epsilon_{\rho\sigma\alpha\beta}
    C_{\mu\nu}^{\phantom{\mu\nu}\alpha\beta}\; ; \; C_{\widetilde{\mu\nu}\rho\sigma} = \frac{1}{2}
    \epsilon_{\mu\nu\alpha\beta} C^{\alpha\beta}_{\phantom{\alpha\beta}\rho\sigma}.
 \end{equation}
 It can be proved the following important properties of the Weyl tensor \cite{art1}:
 \begin{equation}\label{weyld}
    C_{\mu\nu\widetilde{\rho\sigma}} = C_{\widetilde{\rho\sigma}\mu\nu} \;\; ; \; \; C_{\mu\nu\widetilde{\rho\sigma}}
    = C_{\widetilde{\mu\nu}\rho\sigma}.
  \end{equation}
  Skew-symmetry in the first and second pairs of the Weyl tensor indices also enables to view this tensor as an
  operator in $\mathfrak{B}$ \cite{Petrov,Law1,Hervik-Coley},\\
  $$C:\mathfrak{B}\rightarrow\mathfrak{B} \;\; ; \;\; B\mapsto C(B) = T \,,\; \textrm{where} \; \,  T_{\mu\nu} =
  C_{\mu\nu\rho\sigma}B^{\rho\sigma}. $$
  This operator will be called the Weyl operator. It has the important property of sending self-dual bivectors into
  self-dual bivectors and anti-self-dual bivectors into anti-self-dual bivectors. To see this let $Z^\pm_{\mu\nu}$
  pertain to $\mathfrak{D}^\pm$ and $F^\pm\equiv C(Z^\pm)$, then using (\ref{weyld}) and (\ref{bx}) we have:
  \begin{equation}\label{Weyl Z into Z}
    \widetilde{F^\pm}_{\mu\nu} = C_{\widetilde{\mu\nu}\rho\sigma}Z^{\pm\,\rho\sigma}
    =C_{\mu\nu\widetilde{\rho\sigma}}Z^{\pm\,\rho\sigma} = C_{\mu\nu\rho\sigma}\widetilde{Z}^{\pm\,\rho\sigma} =
    \pm\varepsilon C_{\mu\nu\rho\sigma}Z^{\pm\,\rho\sigma} \equiv \pm\varepsilon F^\pm_{\phantom{\pm}\mu\nu}.
  \end{equation}
  So the Weyl operator can be split into a direct sum of two operators, $C = C^+ \oplus C^-$, where $C^+$ is called the   self-dual part of Weyl operator,  and $C^-$ is the anti-self-dual part. $C^+$ sends elements of $\mathfrak{D}^+$   into elements of $\mathfrak{D}^+$ and gives zero when operates in $\mathfrak{D}^-$, while $C^-$ sends elements of   $\mathfrak{D}^-$ into elements of $\mathfrak{D}^-$ and has trivial action in $\mathfrak{D}^+$. Restricting the action of $C^\pm$ to $\mathfrak{D}^\pm$ and using the fact that both operators have vanishing trace we find that $C^\pm$ can have the following algebraic types:\\ \\
  $\left [
  \begin{array}{ll}
    \textbf{Type  O}^{\pm} \rightarrow \;  C^\pm = 0  \\
    \textbf{Type  I}^{\pm} \;\,\rightarrow \; C^\pm $ allows 3 distinct eigenvalues  $  \\
    \textbf{Type  D}^{\pm} \rightarrow  C^\pm $ is diagonalizable with a repeated non-zero eigenvalue   $ \\
    \textbf{Type  II}^{\pm} \rightarrow  C^\pm $ is non-diagonalizable and has a repeated non-zero eigenvalue   $
    \\
   \textbf{Type  III}^{\pm} \rightarrow   \, (C^\pm)^3 = 0\, $and$ \,(C^+)^2 \neq 0 $\,(all eigenvalues are zero)$ \\
    \textbf{Type  N}^{\pm} \rightarrow   \, (C^\pm)^2 = 0 \,$and$\, C^+ \neq 0 $ \,(all eigenvalues are zero)$
  \end{array}
\right.$\\ \\ \\
If operator $C^\pm$ is type I$^\pm$ it is called algebraically general, otherwise it is algebraically special.
An algebraic classification for the full Weyl tensor is made by the composition of the possible types of operators
$C^\pm$. For example, Weyl tensor is said to be of type (I,N) if $C^+$ is type I$^+$ and $C^-$ is type N$^-$. Since
the bundles $\mathfrak{D}^+$ and $\mathfrak{D}^-$ are  interchanged by a simple change of sign in the volume form it
follows that the type (I,N) is intrinsically equivalent to the type (N,I) and so on. At the end there are 21 distinct
algebraic types for the Weyl tensor. If the manifold $(M,g_{\mu\nu})$ is complexified or real with (2,2) signature it
follows that the 21 types are allowed. But if $(M,g_{\mu\nu})$ is a real manifold with Lorentzian or Euclidean
signature the reality condition implies that not all classifications are realizable. In Lorentzian signature the
allowable types are (O,O), (I,I), (D,D), (II,II), (III,III) and (N,N), which are respectively the well known Petrov
types O, I, D, II, III and N, while in Euclidean case the possible algebraic types for the Weyl tensor are (O,O), (O,I),
(O,D),(I,I), (I,D) and (D,D).

This classification to Weyl tensor in complexified manifolds was first obtained in \cite{Plebanski75}, Euclidean case was treated in \cite{Hacyan} while (2,2) signature appeared in \cite{Law2}, in all these references spinor techniques are used. In \cite{Law1} the operator method was used to classify the curvature of  (2,2) signature Einstein manifolds, in such reference fewer types are defined because the vanishing eigenvalues are not distinguished from the non-zero ones. Some general aspects of the operator method to classify tensors in higher-dimensional manifolds were addressed in \cite{Hervik-Coley}. An attempt to classify Weyl tensor in pseudo-Riemannian manifolds in dimensions grater than four was described in \cite{ColeyPSEUD}. More references about the Weyl tensor classification can be found in \cite{art1}.
\section{Finding Principal Null Directions from Eigenbivectors}\label{PNDsection}
 In this section it will be assumed that $(M,g_{\mu\nu})$ is a Lorentzian manifold of dimension four with non-vanishing Weyl tensor. In this kind of
 manifold a real null vector $k^\mu\neq0$, $k^\mu k_\mu=0$, is said to point into a principal null direction(PND) if
  \begin{equation}\label{PND}
  k_{[\alpha}C_{\mu]\nu\rho[\sigma}k_{\beta]}k^\nu k^\rho = 0.
  \end{equation}
 In general a space-time admits four PNDs, but if the Petrov type of the Weyl tensor is special, not type I, then some of these principal null directions coincide and we have less than four independent solutions to equation (\ref{PND}). The Petrov classification can be done entirely in terms of the degeneracy of these directions, which is most easily seen using the spinorial approach due to Penrose \cite{Penrose}. For example, in this approach the Weyl tensor is
 type II if it admits just three independent PNDs, one of which is doubly degenerate \cite{Stephani}. The null vector $k^\mu$ is said to point into a degenerate principal null direction if
 \begin{equation}\label{RPND}
    C_{\mu\nu\rho[\sigma}k_{\beta]}k^\nu k^\rho = 0.
 \end{equation}

The intent of this section is to show that the repeated PNDs are deeply related to the eigenbivectors of the Weyl operator, a result that was implicit in Bel's article \cite{Bel} when he defines the Petrov types, but was not explicitly enunciated and proved. More precisely in this section the following theorem will be proved.\\
  \\
  \textbf{Theorem:} \,\emph{If $Z_{\mu\nu}\neq0$ is a self-dual eigenbivector of the Weyl operator,
  $C_{\mu\nu\rho\sigma}Z^{\rho\sigma} \propto  Z_{\mu\nu}$, and there exists a real vector, $k^\mu\neq0$, such that
  $Z_{\mu\nu}k^\nu = 0 $, then $k^\mu$ points into a repeated PND. Conversely, if $k^\mu$ is a repeated PND then the
  Weyl operator admits a self-dual eigenbivector $Z_{\mu\nu}$ such that $Z_{\mu\nu}k^\nu = 0 $.}
  \\

  Evidently this theorem continues to be valid if instead of self-dual eigenbivectors it is used anti-self-dual eigenbivectors, but to avoid double work only the self-dual case will be treated. To begin let us see that if $C_{\mu\nu\rho\sigma}Z^{\rho\sigma} \propto  Z_{\mu\nu}$ and
  $Z_{\mu\nu}k^\nu = 0$ then $k^\mu$ points in a repeated PND. First note that $k^\mu$ is a null vector:
  \begin{equation}\label{kz3}
  0=iZ_{\mu\nu}k^\nu = \widetilde{Z}_{\mu\nu}k^\nu = \frac{1}{2}k^\nu\epsilon_{\mu\nu}^{\phantom{\mu\nu}\rho\sigma}
  Z_{\rho\sigma}  = \frac{1}{2}\epsilon_\mu^{\phantom{\mu}\nu\rho\sigma}k_{[\nu}Z_{\rho\sigma]}\, \Rightarrow \,
  k_{[\nu}Z_{\rho\sigma]} = 0\,,
  \end{equation}
  contracting this equation with $k^\nu$ and using $Z_{\mu\nu}k^\nu = 0 $ we easily get $k^\nu k_\nu=0$. Now let us
  see that $Z_{\mu\nu}$ must be a simple bivector. Let $e^\mu$ be a vector such that $e^\mu k_\mu = 1$, then
  contracting the right side of equation (\ref{kz3}) with $e^\nu$ we obtain
  \begin{equation}\label{Zsimple}
    Z_{\rho\sigma} =  v_\rho k_\sigma -k_\rho v_\sigma \; \; \textrm{with} \; v_\sigma \equiv Z_{\sigma\nu}e^\nu.
  \end{equation}
If we contract the above equation with $k^{\sigma}$ we get that $v_\sigma k^{\sigma} = 0$. This relation together with (\ref{Zsimple}) produces $v^\rho Z_{\rho\sigma}= (v^\rho v_\rho)k_\sigma$. Now using the self-duality of $Z_{\mu\nu}$ and the definition of $v_{\mu}$ we have,
$$  i(v^\rho v_\rho)k_\sigma =v^\rho \widetilde{Z}_{\rho\sigma} = \frac{1}{2}e_\mu Z^{\rho\mu}
\epsilon_{\rho\sigma\alpha\beta}Z^{\alpha\beta} =\frac{1}{2}e_\mu
\epsilon_{\rho\sigma\alpha\beta}Z^{[\alpha\beta}Z^{\rho]\mu}  $$
But since equation (\ref{Zsimple}) shows that $Z_{\mu \nu}$ is a simple bivector it follows that
$Z^{[\alpha\beta}Z^{\rho]\mu} = 0 $. Thus summarizing we got $k^\mu v_\mu = 0$ and  $v^\mu v_\mu = 0$. Note that
$v^\mu$ must be complex, otherwise we would have $v^\mu\propto k^\mu$ which implies $Z_{\mu\nu} = 0 $. Since $k^\mu$ is real and $v^\mu$ complex it is possible to define a null tetrad  frame for the tangent bundle, $\{l,m,\overline{m},n\}$, such that $l^\mu = k^\mu$, $m^\mu = v^\mu $ and the only non-vanishing contractions between basis vectors are $l^\mu n_\mu = 1 = -m^\mu \overline{m}_\mu$. Using the hypothesis that $Z_{\mu\nu}$ is an
eigenbivector of the Weyl operator together with equation (\ref{Zsimple}) we find:
\begin{equation}\label{cllm}
    0 = l^\nu Z_{\mu\nu} \propto l^\nu C_{\mu\nu\rho\sigma}Z^{\rho\sigma} = 2 l^\nu C_{\mu\nu\rho\sigma}m^\rho
    l^\sigma  \;\;\; \Rightarrow \; C_{\mu\nu\rho\sigma}l^\nu l^\rho m^\sigma = 0
\end{equation}
Using the above equation and its complex conjugate we see that $2m^{(\alpha}\overline{m}^{\sigma)}C_{\mu\nu\rho\sigma}l^\nu l^\rho=0$ . But we can reexpress this tensor equation using the expansion of metric in a null tetrad frame, $g^{\alpha\sigma} = 2l^{(\alpha}n^{\sigma)} - 2m^{(\alpha}\overline{m}^{\sigma)} $:
$$ 0 = C_{\mu\nu\rho\sigma}l^\nu l^\rho[(l^\alpha n^\sigma + n^\alpha l^\sigma) - g^{\alpha\sigma}] \; \Rightarrow \;
l_\alpha C_{\mu\nu\rho\sigma}l^\nu l^\rho  n^\sigma=C_{\mu\nu\rho\alpha}l^\nu l^\rho\,. $$
From which it follows that $k^\mu=l^\mu$ satisfies equation (\ref{RPND}). This means that $k^\mu$ points in a repeated
PND, proving the first part of the theorem.

Now suppose that $k^\mu\neq0$ points in a repeated PND, which means that it obeys (\ref{RPND}). Let us set $l^\mu =
k^\mu $ and complete the null tetrad frame $\{l^\mu,m^\mu,\overline{m}^\mu,n^\mu\}$. Equation (\ref{RPND}) says that
$l_\alpha C_{\mu\nu\rho\sigma}l^\nu l^\rho =l_\mu C_{\alpha\nu\rho\sigma}l^\nu l^\rho $. Contracting this with $m^\mu
n^\alpha$ we get $m^\mu C_{\mu\nu\rho\sigma}l^\nu l^\rho = 0$. Which implies that $T_{\rho\sigma} = C_{\rho\sigma\mu\nu}l^\mu m^\nu$ is a bivector such that $T_{\rho\sigma}l^\sigma = 0$. Now since $\{l^\mu,m^\mu,\overline{m}^\mu,n^\mu\}$ is a basis for the tangent bundle the following expansion is always valid:
$$T_{\mu\nu} = A\,l_{[\mu} m_{\nu]} + B\, l_{[\mu}\overline{m}_{\nu]}+ C\,l_{[\mu}n_{\nu]} + D\,m_{[\mu}\overline{m}_{\nu]} +
E\,  m_{[\mu}\ n_{\nu]}+ F\, \overline{m}_{[\mu}\ n_{\nu]}. $$
Where $A,B,C,D,E$ and $F$ are complex numbers. Using $T_{\rho\sigma}l^\sigma = 0$ it follows that $C=E=F=0$. Also
since the Weyl operator maps self-dual bivectors into self-dual bivectors and $l_{[\mu} m_{\nu]}$ is self-dual then $T_{\mu\nu}$ must also be self-dual. This fact implies that $B=D=0$, so that $T_{\mu\nu} = Al_{[\mu} m_{\nu]}$, \textit{i.e.}, $l_{[\mu}m_{\nu]}$ is a self-dual eigenbivector of the Weyl operator such that $T_{\mu\nu}l^\nu = 0$, finishing the proof of theorem.
\\

This section dealt only with Lorentzian signature, thus natural questions that can be raised are: (1) What should be
the analogues of repeated PNDs in the other signatures and in the case of complexified manifolds? (2) Are these analogues of repeated PNDs related to eigenbivectors of Weyl operator? The above theorem and its proof give some hint toward the answer of these questions, since it was shown that to every repeated principal null direction, $l^\mu$, it is associated a simple self-dual eigenbivector of the Weyl operator, $Z_{\mu\nu} = l_\mu m_\nu - m_\mu l_\nu$, and \textit{vice versa}. So it is natural to guess that the objects that will replace the repeated PNDs in a general four-dimensional manifold are the
simple self-dual eigenbivectors of Weyl operator. Indeed, we will see in forthcoming sections that this conjecture is correct.

\section{Generalization of the Goldberg-Sachs Theorem}\label{GSHP/section}
The most important theorem related to the Weyl tensor classification in Lorentzian manifolds is the Goldberg-Sachs theorem \cite{Goldberg-Sachs}. It states that a vacuum space-time has an algebraically special Weyl tensor if, and only if, the repeated principal null direction generates a null congruence that is geodesic and shear-free. Thirteen years after the proof of this theorem J. Plebla\'{n}ski created a classification to the Weyl tensor in complexified four-dimensional manifolds \cite{Plebanski75}  and proved together with S. Hacyan the analogue of Goldberg-Sachs theorem in these manifolds \cite{Plebanski2}. In this section this important theorem will be trivially generalized to four-dimensional manifolds of all signatures, this generalization is an important ingredient to answer the two questions raised at the end of last section as well as to investigate the geometric consequences of an algebraically special Weyl tensor.

Let $(M,g_{\mu\nu})$ be a four-dimensional manifold (complexified or real of any signature) with vanishing Ricci tensor (vacuum)\footnote{Throughout this and the next section the Ricci tensor will always be assumed to vanish. Also the tangent bundle is assumed to be endowed with a torsion-free connection compatible with the metric (Levi-Civita), only this kind connection is considered in this article.}, so that the Riemann tensor is equal
to the Weyl tensor.  Let $\{e_{1},e_{2},e_{3},e_{4}\}$ be a null tetrad frame for the tangent bundle, defined to be such that the only non-zero contractions are $e_1^\mu e_{3\,\mu} = 1 = -e_2^\mu e_{4\,\mu}$. The components of the metric in this basis are denoted by $g_{ab} = g_{\mu\nu}e_a^\mu e_b^\nu$. The dual
frame of 1-forms is denoted by $\{e^{1},e^{2},e^{3},e^{4}\}$, $e^a(e_b)=\delta^a_b$ \footnote{Note that
$e^{1}_{\phantom{1}\mu} =e_{3\,\mu}$, $e^{2}_{\phantom{2}\mu} =-e_{4\,\mu}$, $e^{3}_{\phantom{3}\mu} =e_{1\,\mu}$ and
$e^{4}_{\phantom{4}\mu} =-e_{2\,\mu}$.}. Let us denote a set of ten components of the Weyl tensor by:
\begin{eqnarray}\label{WeylScalars}
  \Psi^+_0 \equiv C_{1212} \; ; \; \Psi^+_1 \equiv C_{1312} \; ; \; \Psi^+_2 \equiv C_{1243}\; ; \;  \Psi^+_3 \equiv
C_{1343} \; ; \;\Psi^+_4 \equiv C_{3434} \nonumber\\
  \Psi^-_0 \equiv C_{1414} \; ; \; \Psi^-_1 \equiv C_{1314} \; ; \; \Psi^-_2 \equiv C_{1423}\; ; \;  \Psi^-_3 \equiv
C_{1323} \; ; \;\Psi^-_4 \equiv C_{3232}.
\end{eqnarray}
Where, for example, $C_{1312}\equiv C_{\mu\nu\rho\sigma}e_1^{\phantom{1}\mu} e_3^{\phantom{1}\nu}
e_1^{\phantom{1}\rho} e_2^{\phantom{1}\sigma}$ and the scalars $\Psi_A^{\pm}$ are called the Weyl scalars. The self-dual part of the Weyl tensor, $C^+$, depends only on the scalars $\Psi_A^{+}$, while $C^-$ depends only on $\Psi_A^{-}$. The vanishing of the Ricci tensor and the first Bianchi identity satisfied by the Riemann tensor can be summarized by the following equations:
\begin{eqnarray}\label{WeylSimmetriesNulltetrad}
  \nonumber C_{2123} &=& C_{4143}= C_{1214}= C_{3234} = 0 \; ; \\
  C_{2124}  &=& \Psi^+_1\;;\; C_{4142} = \Psi_1^-\;;\;C_{2324} = \Psi_3^-\;;\;C_{4342} = \Psi^+_3\;; \\
  \nonumber C_{2424}&=& C_{1313} =  \Psi^+_2+\Psi_2^-\; ; \; C_{1324} = \Psi_2^--\Psi^+_2.
\end{eqnarray}
The various algebraic types of the Weyl tensor can be characterized by the possibility of annihilating some of the Weyl scalars by a suitable choice of null tetrad \cite{art1}. For example, when $C^+$ is type N$^+$ it is possible to find a null frame in which $\Psi^+_0=\Psi^+_1=\Psi^+_2=\Psi^+_3=0$ and $\Psi^+_4\neq0$. The table below shows which Weyl scalars can be set to zero by conveniently choosing the null tetrad in each of the possible types of $C^+$.
\begin{center}
\textmd{Weyl Scalars that Can be Made to Vanish by a Suitable Choice of Basis}\\
\begin{tabular}{|c|c|c|}

  \hline

  Type O$^+$ - All $\Psi^+_A$ & Type I$^+$ - $\Psi^+_0,\Psi^+_4$ & Type D$^+$ - $\Psi^+_0,\Psi^+_1,\Psi^+_3,\Psi^+_4$  \\ \hline
  Type II$^+$ -$\Psi^+_0,\Psi^+_1,\Psi^+_4$ & Type III$^+$ - $\Psi^+_0,\Psi^+_1,\Psi^+_2,\Psi^+_4 \>$ & Type N$^+$ - $\Psi^+_0,\Psi^+_1,\Psi^+_2,\Psi^+_3$  \\
  \hline
\end{tabular}
\end{center}
Obviously a similar table can be constructed for the types of $C^-$ in terms of $\Psi_A^-$. Particularly note that when $C^+$ is algebraically special, \textit{i.e.} not type I$^+$, it is always possible to find a null frame in which $\Psi^+_0=\Psi^+_1=0$. Conversely, when $\Psi^+_0=\Psi^+_1=0$ in some null tetrad then $C^+$ is algebraically special.

Now choosing conveniently the sign of the volume form it follows that the below bivectors form a basis for the space of self-dual and anti-self-dual bivectors respectively:
 \begin{eqnarray}\label{Zbasis}
  \nonumber Z^{1+} = e^4\wedge e^3 \; ; \; Z^{2+} = e^1\wedge e^2 \; ; \;Z^{3+} = \frac{1}{\sqrt{2}}[e^1\wedge e^3 -
  e^2\wedge e^4] \\
  Z^{1-} = e^2\wedge e^3 \; ; \; Z^{2-} = e^1\wedge e^4 \; ; \;Z^{3-} = \frac{1}{\sqrt{2}}[e^1\wedge e^3 - e^4\wedge
  e^2]
\end{eqnarray}
Note that the bivectors $Z^{3\pm}$ are not simple, since $\widetilde{Z^{3\pm}}^{\mu\nu}Z^{3\pm}_{\mu\nu} =\pm\varepsilon Z^{3\pm\,\mu\nu}Z^{3\pm}_{\mu\nu}=\mp2\varepsilon\neq0$, while $Z^{1\pm}$ and $Z^{2\pm}$ are obviously simple. Now let $Z_{\mu\nu}$ be a simple and self-dual bivector, then it is not difficult to see that it is possible to find a null tetrad frame in which  $Z_{\mu\nu} = Z^{1+}_{\mu\nu} = 2e_{1[\mu}e_{2\nu]}$. Using this frame we have:
$$ C_{\mu\nu\rho\sigma}e_1^{\phantom{1}\rho}e_2^{\phantom{2}\sigma}=\lambda e_{1[\mu}e_{2\nu]}\;\Rightarrow\;\Psi^+_0
= \lambda e_{1\mu}e_{2\nu}e_1^{\phantom{1}[\mu}e_2^{\phantom{2}\nu]}=0\; \; \textrm{and}\;\;\Psi^+_1 = \lambda
e_{1\mu}e_{2\nu}e_1^{\phantom{1}[\mu}e_3^{\phantom{2}\nu]}=0.$$
This means that if the Weyl operator admits a simple self-dual eigenbivector then it is possible to find a null tetrad frame where $\Psi^+_0=\Psi^+_1=0$. Conversely, after some algebra it can be proved that when Weyl operator acts in $Z^{1+}$ we get $C(Z^{1+})=2\Psi^+_2Z^{1+}+2\Psi^+_0Z^{2+}+2\sqrt{2}\Psi^+_1Z^{3+}$, thus if $\Psi^+_0=\Psi^+_1=0$ then $Z^{1+}$ is a simple self-dual eigenbivector of the Weyl operator. Analogously $\Psi^+_3=\Psi^+_4=0$ if, and only if, $Z^{2+}$ is an eigenbivector of the Weyl operator. From preceding results can be stated:\\
\\
\emph{Weyl operator $C^+$ is algebraically special if, and only if, it admits a simple self-dual eigenbivector.
Analogously, $C^-$ is algebraically special if, and only if, it admits a simple anti-self-dual eigenbivector.}\\
\\
Previous section established that in Lorentzian signature to every repeated PND it is associated a simple self-dual
eigenbivector of the Weyl operator. Since the existence of a repeated principal null direction is equivalent to the Weyl tensor being
algebraically special, the above result endorse the conjecture that the natural extension of repeated PNDs to other
signatures are the simple self-dual eigenbivectors of the Weyl operator. The Goldberg-Sachs theorem says that in vacuum space-times the repeated PNDs are tangent to shear-free null geodesic congruences, which is an integrability property related to these directions. So to the conjecture be correct it is important to relate these simple self-dual eigenbivectors of the Weyl operator to some integrability property. This will be accomplished in what follows. \\

Denoting the Levi-Civita connection by $\nabla_\mu$ then the connection 1-forms, $\omega^a_{\phantom{a}b}$, and its
components in the null frame are defined by the following equations:
\begin{equation}\label{connection one forms}
    \nabla_Xe_b \equiv X^\mu\nabla_\mu e_b = \omega^a_{\phantom{a}b}(X)e_a\;\; ; \;\;
    \omega_{ab}^{\phantom{ab}c}\equiv \omega^c_{\phantom{c}b}(e_a)\,,
\end{equation}
where $X$ is an arbitrary vector field. The indices of the connection 1-forms can be lowered by means of the metric,
$\omega_{ab}=g_{ac}\omega^c_{\phantom{c}b}$, $\omega_{abc}=g_{cd}\omega_{ab}^{\phantom{ab}d}$. Since the components of the metric in this frame are constant it follows that $\omega_{ab}=-\omega_{ba}$ and $\omega_{abc}=-\omega_{acb}$. The Cartan structure equations when the Ricci tensor vanishes are given by:
\begin{equation}\label{cartan}
de^a + \omega^a_{\phantom{a}b}\wedge e^b =0\;\, ; \,\; \frac{1}{2}C^a_{\phantom{a}bcd}e^c\wedge e^d =
d\omega^a_{\phantom{a}b} + \omega^a_{\phantom{a}c}\wedge\omega^c_{\phantom{c}b}.
\end{equation}
By means of equations (\ref{WeylScalars}), (\ref{WeylSimmetriesNulltetrad}) and (\ref{Zbasis}) it follows that the
self-dual part of the second structure equation can be written in the below form.
\begin{eqnarray}
  \Psi_2^+\,Z^{1+} + \Psi_0^+\,Z^{2+} + \sqrt{2}\Psi_1^+\,Z^{3+}&=&  d\omega_{12} + \omega_{12}\wedge(\omega_{24}
  -\omega_{13}) \label{structure1} \\
  \Psi_4^+\,Z^{1+} + \Psi_2^+\,Z^{2+} + \sqrt{2}\Psi_3^+\,Z^{3+} &=&  d\omega_{43} - \omega_{43}\wedge(\omega_{24}
  -\omega_{13}) \label{structure2}\\
   -2\Psi_3^+\,Z^{1+} - 2\Psi_1^+\,Z^{2+} - 2\sqrt{2}\Psi_2^+\,Z^{3+} &=&  d(\omega_{24}-\omega_{13}  ) +
   2\omega_{12}\wedge\omega_{43} \label{structure3}
\end{eqnarray}
Making the changes $Z^{i+} \rightarrow Z^{i-}$, $\Psi_A^+\rightarrow\Psi_A^-$,
$\omega_{12}\rightarrow\omega_{14}$, $\omega_{43}\rightarrow\omega_{23}$ and $\omega_{24}\rightarrow\omega_{42}$ in (\ref{structure1}),
(\ref{structure2}) and (\ref{structure3}) we get the other three missing components of the second structure equation.

When the self-dual part of the Weyl tensor vanishes, $\Psi_A^+=0$ for all $A$, it is seen that a possible solution to equations (\ref{structure1}), (\ref{structure2}) and (\ref{structure3}) is $\omega_{12}=\omega_{34}=0$ and $\omega_{24}=\omega_{13}$. Conversely if $\omega_{12}=\omega_{34}=0$ then equations (\ref{structure1}) and (\ref{structure2}) implies that $\Psi_A^+=0$ for all $A$. Since this result will be important in the next section let us stress it:\\
\\
\emph{When $C^+$ vanishes there exists some null frame in which $\omega_{12}=\omega_{34}=0$ and $\omega_{24}=\omega_{13}$. Conversely, if $\omega_{12}=\omega_{34}=0$ in some null frame then the self-dual part of the Weyl tensor is zero. }\\

Now the generalization of the Goldberg-Sachs(GS) theorem to all four-dimensional manifolds will be enunciated.
In what follows this theorem will be dubbed the GSHP theorem, since Pleba\'{n}ski and Hacyan were responsible for the
extension of GS theorem to the case of complexified manifolds of complex dimension four. Here the theorem will be extended in a trivial way to real four-dimensional manifolds of all signatures but the proof will be omitted because it is basically the same of the complexified case \cite{Plebanski2}. \\
\\
\textbf{GSHP Theorem:}  \textsl{Let $(M,g_{\mu\nu})$ be a four-dimensional manifold (complexified or real with any
signature) with vanishing Ricci tensor, then the Weyl scalars $\Psi_0^+$ and $\Psi_1^+$ vanish if, and only if, there
exists a null tetrad frame in which the connection components $\omega_{112}$ and $\omega_{221}$ are both zero}.
\\
\\
Next section will be devoted to explore the consequences of this theorem in all signatures as well as in complexified manifolds. Before this, it is worth mentioning that the GSHP theorem obviously has an analogous version related to the anti-self-dual part of the Weyl tensor. More precisely can be stated that the Weyl scalars $\Psi_0^-$ and $\Psi_1^-$ vanish if, and only if, there exists a null tetrad frame in which the connection components $\omega_{114}$ and $\omega_{441}$ are zero. Also making the changes $e_1\leftrightarrow e_3$ and $e_2\leftrightarrow e_4$ we get that $\Psi_4^+$ and $\Psi_3^+$ vanish if, and only if, there is a null tetrad frame in which the connection components $\omega_{334}$ and $\omega_{443}$ vanish, similarly $\Psi_4^-$ and $\Psi_3^-$ vanish if, and only if, there is a null tetrad frame in which $\omega_{332}$ and $\omega_{223}$ are both zero.

As a last comment it shall be mentioned that the GSHP theorem is also valid if instead of vanishing Ricci tensor it is assumed that the Ricci tensor is proportional to the metric, an Einstein manifold \cite{Plebanski2}. Recently it was investigated in \cite{Nurwoski2} whether a less restrictive condition can be imposed to the Ricci tensor while keeping the GSHP theorem valid. In Lorentzian signature a conformally invariant version of the GS theorem was proved in reference \cite{GS-CottonYork}.

\section{The Consequences of GSHP Theorem }\label{consequences/section}
Let us calculate the Lie bracket of vectors $e_a$ and $e_b$:
$$[e_a,e_b]=e_a^{\phantom{a}\mu}\nabla_\mu e_b-e_b^{\phantom{b}\mu}\nabla_\mu
e_a=\nabla_ae_b-\nabla_be_a=(\omega_{ab}^{\phantom{ab}c}-\omega_{ba}^{\phantom{ab}c})e_c\,. $$
So, using the identity $\omega_{abc}=-\omega_{acb}$, the above relation implies that
\begin{equation}\label{[e1,e2]}
    [e_1,e_2]= (\omega_{12}^{\phantom{12}c}-\omega_{21}^{\phantom{12}c})e_c =
    (\omega_{123}-\omega_{213})e_1-(\omega_{124}-\omega_{214})e_2+ \omega_{121}e_3 + \omega_{212}e_4.
\end{equation}
Since $[e_1,e_1]$ and $[e_2,e_2]$ are trivially zero, it follows from (\ref{[e1,e2]}) that the distribution generated by the vector fields $\{e_1,e_2\}$ is integrable if, and only if, $\omega_{112}=\omega_{221}=0$. Thus what the GSHP theorem says is that the integrability of the planes generated by $\{e_1,e_2\}$ is equivalent to the vanishing of the Weyl scalars $\Psi^+_0$ and $\Psi^+_1$. Since $e_{1}^{\phantom{1}\mu}e_{1\,\mu}=e_{2}^{\phantom{2}\mu}e_{2\,\mu}=
e_{1}^{\phantom{1}\mu}e_{2\,\mu}=0 $, then all vectors tangent to the planes generated by $\{e_1,e_2\}$ are null, this kind of distribution is called isotropic or totally null. More about isotropic subspaces can be found in \cite{Simple Spinors}. Thus, in other words, the theorem proved in the last section states that in a Ricci-flat four-dimensional manifold the Weyl tensor is algebraically special if, and only if, the manifold admits an integrable foliation of isotropic planes. In complexified manifolds this result was obtained in \cite{Plebanski2}, where it was also proved that this two-dimensional foliation is extremal, in the sense that it can be obtained from the extremization of some functional.

Simple bivectors that generate isotropic distributions are called null bivectors. For example, the bivector $Z^{1+}_{\mu\nu} = 2e_{1[\mu}e_{2\nu]}$ is null. In four dimensions a bivector is null if, and only if, it is simple and self-dual or simple and anti-self-dual. In the last section it was demonstrated that the Weyl tensor admits a null eigenbivector if, and only if, there exists some null frame in which $\Psi^+_0=\Psi^+_1=0$. In this frame $Z^{1+}$ is an eigenbivector of the Weyl tensor. But from the above results the distribution generated by $Z^{1+}$, $\{e_1,e_2\}$, is integrable. Thus arriving at the following important result:\\
\\
\emph{In a Ricci-flat manifold the Weyl tensor admits a null eigenbivector if, and only if, the isotropic distribution generated by this bivector is integrable.} \\
\\
The author is not aware of any previous literature that arrived at this statement. Such result shows that the bivector approach to the classification of the Weyl tensor is useful and fruitful to analyze the integrability properties in an unified way in all signatures. So the bivector method used by  A. Z. Petrov \cite{Petrov} and for long time abandoned is, after all, suitable and convenient for some types of studies. Before proceeding it will be introduced some definitions and notation that will be important in what follows.

Given a manifold $(M,g_{\mu\nu})$, an almost complex structure on this manifold is an endomorphism of the tangent bundle, $J:TM\rightarrow TM$, such that its square is minus the identity map, $J(J(V))=-V$ for all $V\in TM$. Let us define the following almost complex structure:
\begin{equation}\label{Jdefinition}
    J\equiv i(e_1\otimes e^1 + e_2\otimes e^2)-i(e_3\otimes e^3 + e_4\otimes e^4)\,.
\end{equation}
It is easy to see that this almost complex structure has the important property of leaving the metric invariant, $g(X,Y)=g(J(X),J(Y)) $ for all $X,Y\in TM$. Because of this the metric is said to be Hermitian with respect to $J$. The operator $J$ naturally splits the tangent bundle into a direct sum
of two bundles of the same dimension,
$$TM= TM^+ \oplus TM^- \;\;\; \textrm{with} \;\;\; TM^\pm\equiv\{V\in TM\, | \,J(V)=\pm iV\} .$$
When both bundles $TM^+$ and $TM^-$ are integrable the almost complex structure $J$ is said to be integrable. For the $J$ defined on equation (\ref{Jdefinition}) we have $TM^+=\textrm{Span}\{e_1,e_2\}$ and $TM^-=\textrm{Span}\{e_3,e_4\}$, so that from the previous results we conclude that $J$ is integrable if, and only if, $\omega_{112}$, $\omega_{221}$, $\omega_{334}$ and $\omega_{443}$ all vanish. It is easy to prove that the integrability of $J$ is equivalent to the vanishing of the Nijenhuis tensor, $N$, defined by the below equation \cite{Nakahara}.
$$N:TM\times TM\rightarrow TM\;\;\;;\;\;\; N(X,Y)=[X,Y]-[J(X),J(Y)]+J([J(X),Y])+J([X,J(Y)]). $$

The K\"{a}hler form, $\Omega$, is the 2-form constructed from $J$ and $g$ whose action on $TM\times TM$ is defined by
$\Omega(X,Y)=g(J(X),Y)$. For the $J$ defined on equation (\ref{Jdefinition}) we have:
\begin{equation}\label{Kahler form}
    \Omega\,=\, i(e^1\wedge e^3 + e^4\wedge e^2)\, = \, i\sqrt{2}\,Z^{3+}.
\end{equation}
For a complex manifold\footnote{Where by complex manifold it is meant a manifold which over the
complex field can be covered by charts with analytic transition functions.} if the metric is Hermitian with respect to an integrable $J$ and $\Omega$ is a closed form, $d\Omega=0$, then the manifold is called a K\"{a}hler manifold. If besides this the curvature is Ricci-flat, as will be assumed in this section, the manifold is said to be a Calabi-Yau manifold\footnote{Actually a Calabi-Yau manifold is defined to be a K\"{a}hler manifold with vanishing
first Chern class. When the Ricci tensor is zero the first Chern class vanishes trivially. Conversely, it can be proved that a K\"{a}hler manifold with vanishing first Chern class admits a Ricci-flat metric.}. For later convenience let us calculate the exterior derivative of the K\"{a}hler form:
$$d\Omega=i[de^1\wedge e^3 - e^1\wedge de^3 + de^4\wedge e^2 - e^4\wedge de^2 ] = i[-\omega^1_{\phantom{1}a} \wedge
e^a\wedge e^3 + e^1 \wedge \omega^3_{\phantom{3}a} \wedge e^a -\omega^4_{\phantom{4}a} \wedge e^a\wedge e^2 + e^4
\wedge \omega^2_{\phantom{2}a} \wedge e^a ]=$$
\begin{equation}\label{domega}
    =  -2i\omega_{12}\wedge e^1\wedge e^2 + 2i\omega_{34}\wedge e^3\wedge e^4
\end{equation}
Since $\omega_{ab}=\omega_{cba}e^c$ then the above equation implies that the closeness of $\Omega$ together with the integrability of $J$ (K\"{a}hler condition) is equivalent to the vanishing of $\omega_{12}$ and $\omega_{34}$. Using the relations $\nabla_ae_b=\omega_{ab}^{\phantom{ab}c}e_c$ and
$\nabla_ae^b=\omega_{a\phantom{b}c}^{\phantom{a}b}e^c$ it is straightforward to calculate $\nabla_a J$. For example, $\nabla_1 J=2\omega_{134}(e_1\otimes e^4+ e_2\otimes e^3)+ 2\omega_{121}(e_3\otimes e^2+ e_4\otimes e^1)$. Computing the other terms we get that $J$ is covariantly constant if, and only if, the connection 1-forms $\omega_{12}$ and $\omega_{34}$ vanish. Then can be stated:
\begin{equation}\label{Jconstant}
J\;\textrm{integrable and} \;\;d\Omega=0 \;\;\;\Leftrightarrow \;\;\; \nabla_XJ=0\;\; \forall\;X\in TM \;\;\;\Leftrightarrow \;\;\;\omega_{12}\,=\,\omega_{34} \,=\,0 \,.
\end{equation}

 Now the consequences of the above results will be investigated in complexified manifolds as well as in real manifolds with all types of signature. Important attempts on the lines presented below can be found in \cite{Robinson Manifolds,Nurwoski2}, where the integrable isotropic structures are investigated in detail. In what follows new results are presented, in particular it is proved that when $C^+$ vanishes the manifold admits a covariantly constant rank two tensor and a solution for the source-free Einstein-Maxwell equations.

\subsection{Complexified Manifolds}\label{complexmanifolds}
In the case of $(M,g_{\mu\nu})$ being a complexified Riemannian manifold of complex dimension four, the ten complex Weyl scalars, $\Psi_A^\pm$, are independent of each other and all the 21 types of classification for the Weyl tensor are realizable. By means of the canonical forms of each type, described in reference \cite{art1}, we can see that if the type of the Weyl tensor is (I,I) then the Weyl operator does not have any null eigenbivector, so that the manifold does not admit integrable isotropic planes. For types (II,I),(III,I) and (N,I) the Weyl tensor admits just one independent null eigenbivector, thus just one foliation of isotropic planes. In types (II,II), (II,III), (II,N), (III,III), (III,N), (N,N) and (D,I) there are two independent distributions of integrable isotropic planes, while types (D,II), (D,III) and (D,N) allow three such distributions. Type (D,D) admits four independent integrable isotropic planes. For types (O,I), (O,II),
(O,D), (O,III), (O,N) and (O,O), when $C^+$ or $C^-$ vanishes, there are infinitely many integrable isotropic planes.

The most interesting results appear when the Weyl tensor is type (D,something) or type (O,something). In these cases there is some null tetrad frame in which $\Psi_0^+=\Psi_1^+=\Psi_3^+=\Psi_4^+=0$ \cite{art1}. As mentioned at the beginning of section \ref{GSHP/section} this implies that $Z^{1+}_{\mu\nu} = 2e_{1[\mu}e_{2\nu]}$ and $Z^{2+}_{\mu\nu} = 2e_{4[\mu}e_{3\nu]}$ are simple self-dual eigenbivectors of the Weyl tensor, so that, as seen above, the planes generated by $\{e_1,e_2\}$ and $\{e_4,e_3\}$ are integrable. But these planes are the ones that generate $TM^+$ and $TM^-$ respectively, thus these bundles are integrable. Conversely suppose that the Ricci-flat manifold $(M,g)$ has an integrable almost complex
structure, $J$. This means that $TM^+$ and $TM^-$ are integrable. If $g$ is a Hermitian metric with respect to $J$ we
have that these bundles are isotropic. For example, if $X^+,Y^+\in TM^+ $ then $g(X^+,Y^+) = g(J(X^+),J(Y^+)) =
g(iX^+,iY^+) = -g(X^+,Y^+) $, so $g(X^+,Y^+) = 0$. Now if $\{e_1,e_2\}$ is some basis of $TM^+$ then, since the metric
is non-degenerate, it is always possible to find a basis $\{E_3,E_4\}$ for $TM^-$ such that
$g(e_1,E_3)=-g(e_2,E_4)=1$, $g(e_1,E_4)=a$, $g(e_2,E_3)=b$. If $a=b=0$ put $e_3=E_3$ and $e_4=E_4$, if $a\neq0$ and
$b=0$ put $e_3=E_3$ and $e_4=E_4-aE_3$, and if $a\neq0\neq b$ and $(1+ba)\neq0$ put $e_3=\frac{1}{1+ba}(E_3+bE_4)$ and
$e_4=\frac{1}{1+ba}(E_4-aE_3)$. The case $a\neq0\neq b$ and $(1+ba)=0$ is not possible, since in this case there would
be an isotropic space of dimension three. Then the vectors $\{e_{1},e_{2},e_{3},e_{4}\}$ form a null tetrad frame and the planes generated by  $\{e_{1},e_{2}\}$ and $\{e_{3},e_{4}\}$ are integrable, this implies that $\Psi_0^+=\Psi_1^+=0$
and $ \Psi_3^+=\Psi_4^+=0$, so Weyl tensor is type (D,something) or type (O,something). This paragraph can be
summarized by the following words:\\
\\
\emph{In a Ricci-flat complexified manifold the self-dual part of the Weyl operator, $C^+$, is type D$^+$ or type O$^+$ if, and only if, there is some null tetrad frame in which the almost complex structure defined in (\ref{Jdefinition}) is integrable. In other words, $C^+$ is type D$^+$ or type O$^+$ if, and only if, the Ricci-flat complexified manifold admits an integrable almost complex structure such that the metric is Hermitian with respect to it. }
\\

In the particular case of Weyl tensor being type (O,something), $\Psi_A^+=0$ for all $A\in (0,1,2,3,4)$, equations (\ref{structure1}), (\ref{structure2}) and (\ref{structure3}) implies that there is some null tetrad frame where the connection 1-forms $\omega_{12}$ and $\omega_{43}$ vanish while $\omega_{24}=\omega_{13}$. This together with equation (\ref{domega}) implies that $J$ is integrable and the exterior derivative of the K\"{a}hler form is zero\footnote{Note that if $\omega_{12}=0$, $\omega_{43}=0$ and $\omega_{24}=\omega_{13}$ then all the isotropic distributions $\{ae_1+be_4,ae_2+be_3\}$ for $a,b$ constants are integrable. Thus anti-self-dual manifolds admit infinitely many integrable self-dual isotropic distributions.}. The Euclidean version of this result was previously obtained in \cite{Broda}. Now according to (\ref{Jconstant}) this is equivalent to $J$ being a constant tensor. Conversely, if $J$ is an integrable almost complex structure and $d\Omega=0$, then equation (\ref{domega}) together with the reasoning of the last paragraph implies that there exists a null tetrad frame such that $\{e_1,e_2\}$ generates $TM^+$ and $\{e_3,e_4\}$ generates $TM^-$ and such that $\omega_{12}=\omega_{34}=0$. Inserting this last equality into equations (\ref{structure1}) and (\ref{structure2}) we have $\Psi_0^+=\Psi_1^+=\Psi_2^+=\Psi_3^+=\Psi_4^+=0$. Thus we can state:\\
\\
\emph{In a Ricci-flat complexified manifold the self-dual (or the anti-self-dual) part of the Weyl operator, $C^+$ ($C^-$),  vanishes if, and only if, there is some null tetrad frame in which the almost complex structure defined in (\ref{Jdefinition}) is covariantly constant. In the case of a complex manifold this means that $C^+$ (or $C^-$) vanishes if, and only if, the manifold is Calabi-Yau.}

\subsection{Euclidean Signature}\label{Euclidean/subsection}
In a real four-dimensional Euclidean Ricci-flat manifold $(M,g)$, given an orthonormal frame,
$\{E_{1},E_{2},E_{3},E_{4}\}$ with $g(E_a,E_b)=\delta_{ab}$, it is possible to construct the following null tetrad
frame in the complexified tangent bundle, $\mathbb{C}\otimes TM$: $e_1=\frac{1}{\sqrt{2}}(E_{1}+iE_{2}) \, , \,
e_3=\frac{1}{\sqrt{2}}(E_{1}-iE_{2})\,,\, e_2=\frac{1}{\sqrt{2}}(E_{3}+iE_{4}) \; \textrm{and}
\;e_4=\frac{-1}{\sqrt{2}}(E_{3}-iE_{4})\,$. Note that since $\{E_a\}$ are real vector fields it follows that
$\overline{e_1}=e_3$ and $\overline{e_2}=-e_4$. For the purposes of this section it can always be assumed that the basis vectors obey to these reality conditions. Therefore the almost complex structure defined in
(\ref{Jdefinition}) and the K\"{a}hler form of equation (\ref{Kahler form}) are real tensors, $\overline{J}=J$ and
$\overline{\Omega}=\Omega$, while the Weyl scalars are such that $\overline{\Psi^{\pm}_0} = \Psi^{\pm}_4$ ,
$\overline{\Psi^{\pm}_1} = -\Psi^{\pm}_3$ and $\overline{\Psi^{\pm}_2} = \Psi^{\pm}_2$. This implies that the only
allowed types for the Weyl tensor are (O,O), (O,I), (O,D), (I,I), (I,D) and (D,D) \cite{art1}. In particular note that if the Weyl tensor is algebraically special, not type (I,I), then conveniently choosing the sign of the volume form we can guarantee that $C^+$ is type D$^+$ or type O$^+$.

Results of subsection \ref{complexmanifolds} implies that if $C^+$ is type D$^+$ or O$^+$ there is some null frame such that the almost complex structure defined in (\ref{Jdefinition}) is integrable. If $C^+$ is strictly type O$^+$ the K\"{a}hler form is closed and the almost complex structure $J$ is covariantly constant. Since the real 2-form $\Omega$ is non-degenerate it follows that when the self-dual part of the Weyl tensor vanishes, $C^+$ is type O$^+$, the real manifold is symplectic, with symplectic form $\Omega$.

An important theorem in complex differential geometry \cite{Newlander}, the Newlander-Nirenberg theorem, states that a manifold admits an integrable and real almost complex structure if, and only if, it is a complex manifold\footnote{Meaning that the manifold over the complex field can be covered by charts with analytic transition functions.}. Since in Euclidean case $J$ is real it follows that when $C^+$ is type D$^+$
or type O$^+$ the manifold over the complex field is a complex manifold. In the particular case of $C^+$ being type O$^+$ the
manifold is a K\"{a}hler manifold, more precisely a Calabi-Yau, since Ricci tensor is assumed to vanish. Conversely, as seen in last subsection, if the manifold is Calabi-Yau then the self-dual part of the Weyl tensor must vanish. However it must be noted that a manifold can be Calabi-Yau but constructed from the complexification of a non-Euclidean real manifold. The above results are summarized by the following stressed results, one of which is here dubbed the Euclidean version of the GS theorem:
\\
\\
\emph{When the Weyl tensor in a Ricci-flat Euclidean manifold, $M$, is not type (I,I) the manifold over the complex
field is a complex manifold. Particularly, if the self-dual part of the Weyl tensor vanishes then the complexification of
$M$ is a Calabi-Yau manifold and the real tensor $J$ is covariantly constant. Conversely, when the manifold is
Calabi-Yau the self-dual part of the Weyl tensor must vanish, although the manifold may not be the complexification of
a real Euclidean manifold. }
\\
\\
\textbf{Euclidean version of GS theorem:} \emph{In vacuum the Weyl tensor is algebraically special if, and only if,
the tangent bundle admits a real integrable almost complex structure.}  \\
\subsection{Lorentzian Signature}
Lorentzian four-dimensional manifolds are characterized by the existence of a real frame \{$e_t,e_x,e_y,e_z$\} such
that the only non-zero contractions are $e^\mu_t e_{t\mu} = 1$ and $e^\mu_x e_{x\mu} = e^\mu_y e_{y\mu} = e^\mu_z
e_{z\mu} =-1$. Null tetrad frames can be constructed in the complexification of the tangent bundle, one example being $e_1=l=\frac{1}{\sqrt{2}}(e_t+e_z)$, $e_2=m=\frac{1}{\sqrt{2}}(e_x+ie_y)$, $e_3=n=\frac{1}{\sqrt{2}}(e_t-e_z)$  and $e_4=\overline{m}=\frac{1}{\sqrt{2}}(e_x-ie_y)$. Note that vector fields $e_1$ and $e_3$ are real while $e_2$ and $e_4$ are complex and conjugates to each other, therefore the Weyl scalars $\Psi_A^-$ are the complex conjugates of $\Psi_A^+$ and only types (O,O), (I,I), (D,D), (II,II), (III,III) and (N,N) are realizable in this signature \cite{art1}.

When the space-time is algebraically special there is some null tetrad frame in which $\Psi_0^+=\Psi_1^+=\Psi_0^-=\Psi_1^-=0$. From the GSHP theorem it follows that for a vacuum space-time in this frame we have $\omega_{112}=\omega_{221}=\omega_{114}=\omega_{441}=0$, so that:
$$\nabla_1e_1=\omega_{11}^{\phantom{11}a}\,e_a=\omega_{113}\,e_1-\omega_{114}\,e_2+\omega_{111}\,e_3-\omega_{112}\,
e_4=\omega_{113}\,e_1.$$
This means that $e_1$ is a null geodesic vector field. Now let us calculate the optical scalars of the null congruence generated by $e_1$. To accomplish this we must compute the projection of the tensor $B_\mu^{\phantom{\mu}\nu} =
(\nabla_\mu e_1)^\nu$ into the space-like plane generated by $e_x=\frac{1}{\sqrt{2}}(e_2+e_4)$ and
$e_y=\frac{1}{i\sqrt{2}}(e_2-e_4)$ \cite{Wald}.
\begin{multline*}
  \nabla_xe_1= \frac{1}{\sqrt{2}}(\nabla_2e_1+\nabla_4e_1) \sim -\frac{1}{\sqrt{2}}[ (\omega_{214}+\omega_{414})e_2
+(\omega_{212}+\omega_{412})e_4] = -\frac{1}{2}[(\omega_{214}+\omega_{412})e_x+i(\omega_{214}-\omega_{412})e_y ] \\
 \nabla_ye_1= \frac{1}{i\sqrt{2}}(\nabla_2e_1-\nabla_4e_1) \sim -\frac{1}{i\sqrt{2}}[ (\omega_{214}-\omega_{414})e_2
+(\omega_{212}-\omega_{412})e_4] = -\frac{1}{2i}[(\omega_{214}-\omega_{412})e_x + i(\omega_{214}+\omega_{412})e_y ]
\end{multline*}
Where the symbol $\sim$ means equal except for terms proportional to $e_1$. From this equation we conclude that the
projection of $B_\mu^{\phantom{\mu}\nu}$ into the plane $\{e_x,e_y\}$, $\widehat{B}_\mu^{\phantom{\mu}\nu}$, is:
$$
\widehat{B}_\mu^{\phantom{\mu}\nu} = \left[
  \begin{array}{cc}
    B_x^{\phantom{x}x} & B_x^{\phantom{x}y} \\
    B_y^{\phantom{y}x} & B_y^{\phantom{y}y} \\
  \end{array}
\right] = -\frac{1}{2}\left[
  \begin{array}{cc}
    (\omega_{214}+\omega_{412}) & i(\omega_{214}-\omega_{412}) \\
    -i(\omega_{214}-\omega_{412}) & (\omega_{214}+\omega_{412}) \\
  \end{array}
\right] $$
Since the trace-less symmetric part of the above matrix is zero we conclude that the congruence generated by $e_1$ is
shear-free\footnote{In the Newman-Penrose formalism the shear parameter is given by $\sigma=\omega_{212}$, which is
zero in the considered case.}. The expansion of the congruence is the trace of $\hat{B}$, while the rotation is the
skew-symmetric part of this matrix. From section \ref{PNDsection} we know that $e_1$ is a repeated PND when
$\Psi_0^+=\Psi_1^+=0$, so we arrived at the important result that algebraically special space-times in vacuum allow a
shear-free null geodesic congruence generated by the repeated principal null direction. From above results it is easy
to see that the converse is also true, which proves the usual version of the Goldberg-Sachs theorem. In particular, when Weyl tensor is type D there are two independent repeated PNDs, so two independent shear-free null geodesic congruences, this was the key property that enabled to find all type D vacuum solutions of Einstein's equation \cite{typeD}.
\subsection{(2,2) Signature}
Let $(M,g)$ be a Ricci-flat real manifold of (2,2) signature, then  it is possible to find a real frame
$\{E_{1},E_{2},E_{3},E_{4}\}$ such that the only non-zero inner products between the basis vectors are $E_{1}^\mu
E_{1\mu}=E_{2}^\mu E_{2\mu} = 1$ and  $E_{3}^\mu E_{3\mu}=E_{4}^\mu E_{4\mu}=-1$. From this we can construct the
following null tetrad basis in the complexified tangent bundle: $e_1=\frac{1}{\sqrt{2}}(E_{1}+iE_{2}) \, , \,
e_3=\frac{1}{\sqrt{2}}(E_{1}-iE_{2})\,,\, e_2=\frac{1}{\sqrt{2}}(E_{3}+iE_{4}) \, ,
\,e_4=\frac{1}{\sqrt{2}}(E_{3}-iE_{4})\,$. In this complex frame note that $\overline{e_1}=e_3$ and
$\overline{e_2}=e_4$. But in this signature it is also possible to form a real null frame, one example being:
$\check{e}_1=\frac{1}{\sqrt{2}}(E_{1}+E_{3})$, $\check{e}_3=\frac{1}{\sqrt{2}}(E_{1}-E_{3})$,
$\check{e}_2=\frac{1}{\sqrt{2}}(E_{2}+E_{4})$ and $\check{e}_4=\frac{-1}{\sqrt{2}}(E_{2}-E_{4})$. Using a real null
frame it is easy to see that all the 10 Weyl scalars are real and independent of each other, so that in (2,2) signature spaces all the 21 algebraic types for the Weyl tensor are realizable \cite{art1}.

 When $C^+$ is algebraically special there exists some null tetrad frame in which $\Psi_0^+ = \Psi_1^+=0$, but the frame can be complex, as $\{e_a\}$, or real, as $\{\check{e}_a\}$. Thus the GSHP theorem guarantees that if $C^+$ is not type I$^+$ then there is some null frame, complex or real, in which
$\omega_{112}=\omega_{221}=0$. This implies that the planes generated by $\{e_1,e_2\}$ or
$\{\check{e}_1,\check{e}_2\}$ are integrable, in the former case since  $\overline{\omega}_{112}= \omega_{334}$ and
$\overline{\omega}_{221}= \omega_{443}$, then $\omega_{334}$ and $\omega_{443}$ are also zero, so that the planes generated by $\{e_3,e_4\}$ are integrable too. Thus a generalization of Goldberg-Sachs theorem in (2,2) signature can be as follows. \\
\\
  \textbf{(2,2) signature version of GS theorem:} \emph{In vacuum the Weyl tensor is algebraically special if, and only
  if, there is some integrable isotropic distribution of planes in the manifold. The planes can be complex or real, in
  the former case it follows that the complex conjugate planes are also isotropic and integrable.}\\

In (2,2) signature when dealing with the real null tetrad frames it is useful to introduce the following analogues of
$J$ and $\Omega$:
 \begin{equation}\label{J^}
 \check{J}\equiv (\check{e}_1\otimes \check{e}^1 + \check{e}_2\otimes \check{e}^2)-(\check{e}_3\otimes \check{e}^3 +
 \check{e}_4\otimes \check{e}^4)\;\;\; ;\;\;\;  \check{\Omega}=(\check{e}^1\wedge \check{e}^3 + \check{e}^4\wedge
 \check{e}^2).
 \end{equation}
By equations (\ref{domega}) and (\ref{Jconstant}) we get the relations
 \begin{equation}\label{J^constant}
 d\check{\Omega} =  -2\omega_{12}\wedge \check{e}^1\wedge \check{e}^2 + 2\omega_{34}\wedge \check{e}^3\wedge
 \check{e}^4 \;\;\;;\;\check{J}\;\textrm{integrable  and} \;\; d\check{\Omega}=0 \;\;\;\Leftrightarrow \;\;\;
 \nabla_X\check{J}=0\;\; \forall\;X\in TM\,.
 \end{equation}
Where $\check{J}$ is called integrable when its invariant subspaces are integrable. Note that if $X,Y\in TM$ then
$\check{J}(\check{J}(X))=X$ and $\check{\Omega}(X,Y)=g(\check{J}(X),Y)$. The advantage of $\check{J}$ and
$\check{\Omega}$ is that they are real, since the basis $\{\check{e}_a\}$ is real, just as $J$ and $\Omega$ are real
in a complex frame such that $\overline{e_1}=e_3$ and $\overline{e_2}=e_4$. But it is worth noting that the metric is not invariant under $\check{J}$, instead now we have $g(\check{J}(X),\check{J}(Y))=-g(X,Y)$. The tensor $\check{J}$ is called a paracomplex structure, more about this kind of object in the context of GSHP theorem can be found in \cite{parastruct}.

Now if $C^+$ is type D$^+$ there are two possible cases: (1) The null frame in which $\Psi_0^+ = \Psi_1^+=\Psi_3^+ =
\Psi_4^+=0$ is complex, like $\{e_a\}$, in this case the complex planes $\{e_1,e_2\}$ and $\{e_3,e_4\}$ are
integrable, so that the almost complex structure $J$ is integrable. Since the tensor $J$ is real the
Newlander-Nirenberg theorem guarantees that the manifold over the complex field is a complex manifold; (2) The null
frame in which $\Psi_0^+ = \Psi_1^+=\Psi_3^+ = \Psi_4^+=0$ is real, like $\{\check{e}_a\}$, in this case the real planes
$\{\check{e}_1,\check{e}_2\}$ and $\{\check{e}_3,\check{e}_4\}$ are integrable, so that $\check{J}$ is integrable.

When $C^+$ is type O$^+$ there is some null tetrad frame where the connection 1-forms are such that
$\omega_{12}=0=\omega_{34}$ and $\omega_{24}=\omega_{13}$. This null frame can be complex or real, so that we again
have two cases: (1) If this null frame is complex we have that $J$ is integrable, covariantly constant and $d\Omega=0$. The manifold over the real field is symplectic, since $d\Omega=0$ and $\Omega$ is real and non-degenerate, and over the complex field the manifold is Calabi-Yau; (2) If this null frame is real it follows that $\check{J}$ is integrable and covariantly constant and $d\check{\Omega}=0$, which implies that the manifold is symplectic, since $\check{\Omega}$ is real and non-degenerate. The above results are summarized by the following words:\\
\\
\emph{Let $(M,g)$ be a Ricci-flat real manifold of (2,2) signature, then if the Weyl tensor is type (D,something) or type (O,something) there are two distinct families of isotropic planes which are integrable. When these planes are complex it follows that the manifold over the complex field is a complex manifold. If the Weyl tensor is strictly type (O,something) there is a covariantly constant real tensor of rank two and the manifold is symplectic and if the integrable isotropic planes are complex it follows that over the complex field the manifold is Calabi-Yau.} \\

Conversely, it is easy to see that if a real four-dimensional manifold of (2,2) signature admits a real integrable
almost complex structure such that the metric is Hermitian, then $C^+$ is type D$^+$ or type O$^+$. If this almost
complex structure is covariantly constant it follows that $C^+$ vanishes. Analogously, when a real four-dimensional manifold of (2,2) signature admits an integrable real paracomplex structure, $\check{J}$, then $C^+$ is type D$^+$ or type O$^+$, if the tensor $\check{J}$ is covariantly constant it follows that $C^+$ vanishes.

\subsection{A Solution for Einstein-Maxwell Equations}
The source-free Maxwell's equations can be put in the form $dF=0=d\widetilde{F}$, where $F$ is a 2-form and
$\widetilde{F}$ its Hodge dual. In the presence of the electromagnetic field $F$ Einstein's equation becomes
$R_{\mu\nu}-\frac{1}{2}Rg_{\mu\nu}=T_{\mu\nu}$, where $T_{\mu\nu}=F_{\mu\alpha}F_\nu^{\phantom{\nu}\alpha} -
\frac{1}{4}F_{\alpha\beta}F^{\alpha\beta}g_{\mu\nu}$ is the energy-momentum tensor of the electromagnetic field.

Note that $\Omega$ defined on equation (\ref{Kahler form}) is self-dual, so if $d\Omega=0$ then it also follows that $d\widetilde{\Omega}=0$. It has been proven in the preceding subsections that when the self-dual part of the Weyl tensor vanishes, $C^+=0$, in Ricci-flat manifolds, there is some null frame in which the K\"{a}hler form is closed, $d\Omega=0$, which also implies $d\widetilde{\Omega}=0$. Now putting $F=\Omega$ we have that $F$ obeys to the source-free Maxwell's equations. Furthermore computing the energy-momentum tensor associated to $F$ it is easily found that it vanishes, $T_{\mu\nu}=0$. Since the Ricci tensor is assumed to vanish it follows that Einstein's equation in the presence of this field is satisfied. Thus can be stated: \\

\emph{When $C^+$ vanishes in a Ricci-flat manifold there exists some null tetrad frame such that the
K\"{a}hler form defined in (\ref{Kahler form}) is a solution for the source-free Einstein-Maxwell equations. }
\section{Physical Relevance}\label{Physsection}
Although rather mathematical, the issues treated in this article have multiple physical applicability. In this section it will be briefly mentioned some physical areas where the subjects treated on this work are of importance.

The path followed here was to consider complexified manifolds and when convenient choose some reality condition to obtain a real manifold of the desired signature. This kind of approach to obtain results on real four-dimensional Lorentzian manifolds was taken before, for example, in a series of papers of C. McIntosh and M. Hickman \cite{McIntosh}, where it was advocated that the use of complexified manifolds are profitable to obtain real solutions to Einstein's equation. In particular this approach is important to understand better what happens when a Wick rotation is made.

Besides the ubiquitous Lorentzian case, other signatures are also of relevance in several physical problems. Euclidean manifolds, called gravitational instantons, are useful when time is Wick rotated, with the intent of calculating partition functions or in the computation of tunneling probabilities. Whereas (2,2) signature is physically important in classical mechanics, since phase spaces are manifolds of split signature. The (2,2) signature is also of interest for the theory of integrable systems \cite{Mason}. From the results of this article one with direct physical applicability is the existence of covariantly constant real tensors of rank two when $C^+$ vanish, these constant tensors furnish conserved quantities, useful for the incorporation of symmetries in physical problems and for the integrability of equations of motion.

The physical intuition that causal structure is the key concept in general relativity lead Roger Penrose to introduce the complex null tetrad in general relativity, giving rise to the so called Newman-Penrose formalism. This formalism was used in the solution of many important problems in general relativity, one example being the finding of all type D vacuum solutions \cite{typeD}. A message that can be extracted from this is that null directions are connected to many physically important properties and even when they are complex physical content may be encapsulated. Here the isotropic planes were of central importance, since they are generated by null directions Penrose's intuition is enforced by the results of the present paper.

Higher-dimensional space-times have been intensively investigated for long time. One of the lines of research in this topic is the generalization of Petrov classification and correlated results to dimensions greater than four. A successful generalization of Weyl tensor classification in Lorentzian higher-dimensional manifolds was described in \cite{CMPP}, the so called CMPP classification. Since it is well known that Goldberg-Sachs theorem cannot be trivially generalized to higher dimensions\footnote{For instance, reference \cite{FrolovMyers5D} proved that the repeated PNDs of 5-dimensional Myers-Perry black hole are not shear-free}, a partial generalization of GS theorem is being looked for and some important results have been already obtained. In \cite{Dur-Reall} it was proved that every Einstein space-time that admits a multiple Weyl aligned null direction(WAND)\footnote{In CMPP classification the WANDs  are natural higher-dimensional analogues of the four-dimensional principal null directions \cite{CMPP}.} also admits a multiple WAND that is tangent to a geodesic congruence. Further in \cite{Ortaggio5} it was worked out the restrictions on the optical matrix of null congruences tangent to geodesic multiple WAND in five-dimensional Einstein space-times. The present article stressed the importance of null structures (isotropic planes) in four dimensions. Such line of thinking can be used for a higher-dimensional generalization of Petrov classification and GS theorem. This path was followed in \cite{HigherGSisotropic1,HigherGSisotropic2}, where the integrability condition of maximally isotropic hyper-planes is related to algebraic conditions on the Weyl and Cotton-York tensors, which can be seen as the generalization of half of the GS theorem. As a last comment it is worth remembering that maximally isotropic subspaces are associated with the so called pure spinors, a mathematical object with increasing relevance in physical theories, string theory being an example \cite{Nathan}.

\section{Conclusion}
It is well known that in vacuum Lorentzian manifolds the repeated principal null directions of the Weyl tensor are related to the integrability of shear-free null geodesic congruences. It then follows a connection between algebraic constraints on the Weyl tensor and geometrical properties of space-time. Here it has been shown that the same kind of connection happens in Ricci-flat four-dimensional complexified manifolds as well as in real manifolds with any signature. The main results presented in this article
were:\begin{itemize}
                                    \item The analogues of repeated PNDs in non-Lorentzian manifolds are the null eigenbivectors of the Weyl operator.
                                    \item When the Ricci tensor vanishes these eigenbivectors generate integrable
                                        isotropic planes, this is the generalization of the Goldberg-Sachs theorem to
                                        non-Lorentzian manifolds.
                                          \item In Ricci-flat complex manifolds the self-dual part of the Weyl tensor vanishes if, and only if, the manifold is Calabi-Yau.
                                    \item In Ricci-flat Euclidean manifolds the Weyl tensor is algebraically special
                                        if, and only if, the manifold has an integrable almost complex structure and
                                        the metric is Hermitian with respect to it. When the self-dual part of the
                                        Weyl tensor vanishes there is a real covariantly constant rank two tensor and
                                        the manifold over the complex field is Calabi-Yau.
                                    \item In a Ricci-flat (2,2) signature manifold if the self-dual part of the Weyl
                                        tensor vanishes then the manifold is symplectic and has a real covariantly
                                        constant rank two tensor.
                                    \item In all Ricci-flat manifolds such that the self-dual part of the Weyl tensor
                                        vanishes the K\"{a}hler form is a solution to the source-free Einstein-Maxwell
                                        equations.

                                  \end{itemize}
\section*{Acknowledgments}
I want to thank Bruno G. Carneiro da Cunha for the encouragement and for the manuscript revision. This research was
supported by CNPq(Conselho Nacional de Desenvolvimento Cient\'{\i}fico e Tecnol\'{o}gico). The final publication is available at \textit{link.springer.com} 
(DOI:10.1007/s10714-013-1539-4).

\end{document}